\documentstyle[prl,aps,multicol, axodraw,epsf]{revtex}
\tighten
\begin{document}
\draft
\title{
Tunneling Anomaly in Superconductor above Paramagnetic Limit
}
\author{I.L. Aleiner$^{1}$ and B.L. Altshuler$^{1,2}$}
\address{$^{1}$NEC Research Institute,
4 Independence Way, Princeton, NJ 08540\\ 
$^{2}$
 Physics Department, Princeton University, Princeton, NJ 08544
}
\maketitle

\begin{abstract}
We study the tunneling density of states (DoS) in the superconducting
systems driven by Zeeman splitting $E_Z$ into the paramagnetic
phase. We show that, even though the BCS gap disappears,
superconducting fluctuations cause a strong DoS singularity in the
vicinity of energies $-E^\ast$ for electrons polarized along the
magnetic field and $E^\ast$ for the opposite polarization.  The
position of the singularity $E^\ast=\case{1}{2}\left(E_Z +
\sqrt{E_Z^2- \Delta^2} \right)$ (where $\Delta$ is BCS gap at $E_Z=0$)
 is universal. We found analytically the
shape of the DoS for different dimensionality of the
system. For ultrasmall grains the singularity has the form of the hard
gap, while in higher dimensions it appears as a significant though
finite dip.  Our results are consistent with recent experiments in
superconducting films.

\end{abstract}

\begin{multicols}{2}

It is  known that magnetic field, $H$, suppresses
superconductivity because it violates the time reversal
symmetry\cite{Textbook}. Typically, this symmetry breaking is
associated with effect of the magnetic field on the orbital motion of
electrons. However, in some physical situations, the main mechanism of the
destruction of the Cooper pairing is the Zeeman splitting of states
with the same spatial wave-functions but opposite spin directions. One
can consider, {\em e.g.}, a thin superconducting film placed in
magnetic field parallel to the plane of the
film\cite{Adams}. Recently, another possibility was exploited
experimentally\cite{Ralph} -- ultrasmall superconducting grains. In
both cases, the size of a Cooper pair is restricted geometrically and
a flux through this pair reaches flux quantum at fields
higher that the bulk critical filed $H_{c_2}$.  We assume that
superconductivity is already destroyed  by the Zeeman splitting $E_Z =
g_L\mu_BH$ when it happens\cite{footnote}, (here $g_L$ is
the $g$-factor Lande, and $\mu_B$ is the Bohr magneton).

Strictly speaking, the spin splitting destroys superconductivity as
soon as\cite{Larkin64} $E_Z\geq \sqrt{2}\Delta$, with $\Delta$ being
the superconducting gap. This transition from superconducting to
paramagnetic normal state is of the first order, and these phases
coexist in the interval $\Delta\leq E_Z \leq 2\Delta$. From now on, we
assume the condition $E_Z > \sqrt{2}\Delta$.

One might expect that the Cooper pairing is irrelevant for the
properties of the normal paramagnetic phase. In this Letter we show
that, on the contrary, there are clear and observable effects of the
pairing in paramagnetic state {\em even far from the transition region}.
 
One of the most fundamental manifestation of the superconductivity is the
gap in the tunneling density of states (DoS) around the zero energy\cite{Giaver}.
This gap apparently disappears when the system becomes
paramagnetic. We will show that at the same time there appears
a dip in the DoS. The shape of this dip
depends on the dimensionality of the system. However, its position
$E^\ast$ is remarkably universal:
\begin{equation}
E^\ast= \frac{1}{2}\left(E_Z + \sqrt{E_Z^2-\Delta^2}\right)
\label{position}
\end{equation}
for $OD$ (grain), $1D$ (strip) and $2D$ (film) cases.

This result should be compared with the the singularity in the
DoS due to the usual superconducting fluctuations in the
normal metal. According to Ref.~\cite{Altshuler85}, in the magnetic
field another anomaly appears in addition
to the zero bias anomaly when bias $V$ corresponding to Zeeman
splitting $eV_s = E_Z$. The singularity we consider here is positioned
at substantially lower energy and, as we will see, is 
stronger than those considered in Ref.~\cite{Altshuler85}.

Recently, singularity of this type was observed in granular $Al$ film
in a parallel magnetic field\cite{Adams}. We believe that the
deviation from the law $eV_s=E_Z$ observed in Ref.~\cite{Adams} is
coherently explained by our Eq.~(\ref{position}).

We begin with the simplest but instructive case of $0D$ system
(ultrasmall grain). The Hamiltonian $\hat{\cal H}$ of this system can
be written in the basis of the exact single-electron states for
non-interacting system as
\begin{equation}
\hat{\cal H}=\sum_{i;\ \sigma = \uparrow,\downarrow }
{E}_{i,\sigma}
a^\dagger_{i,\sigma}a_{i,\sigma}
- \lambda\bar{\delta} \sum_{i,j}
a^\dagger_{i,\uparrow}
a^\dagger_{i,\downarrow}
a_{j,\downarrow}
a_{j,\uparrow}.
\label{0D}
\end{equation}
Indices $i, j$ and $\sigma = \uparrow,\downarrow$ label the orbital
and spin state of an electron: its total energy $E_{i,\sigma}$ is the
sum of orbital ${\epsilon_i}$ and spin parts, ${E}_{i,\uparrow
(\downarrow)}=\epsilon_i\mp E_Z/2$; and $a^\dagger_{i,\sigma},
a_{i,\sigma}$ are the corresponding fermionic creation-annihilation
operators. Finally, $\bar{\delta}=
\langle\epsilon_{i+1}-\epsilon_i\rangle$ is the mean level spacing and
$\lambda$ is the dimensionless interaction constant.  In
Eq.~(\ref{0D}), we omitted some diagonal terms (those with orbital indices
equal pariwise), which with the help of Hartree-Fock approximation 
can be included into the definition of the eigenenergies
$\epsilon_i$. We also omitted off-diagonal terms 
(involving the fermionic
operators with non-paired indices): the corresponding matrix elements
are known\cite{1overg} to be smaller than the diagonal ones  by a
large factor $1/g$, where $g\gg 1$ is the dimensionless conductance of
the grain. Hamiltonian
(\ref{0D}) is nothing but the usual BCS Hamiltonian, and it was used
 in numerous publications\cite{publications} on the
properties of ultrasmall superconducting grains.

BCS instability which system (\ref{0D}) has at temperature $T=0$ and
$H=0$, disappears as soon as $E_Z$ exceeds $\sqrt{2}\Delta$. In the
absence of the superconducting gap, the structure of the ground state
is similar to that without interaction: orbitals with energies
$|{\epsilon}_i| > E_Z/2$ are spinless (orbitals with $\epsilon_i <
-E_Z/2$ are double ocupied while those with  $\epsilon_i > E_Z/2$ are
empty), while orbitals in the energy strip $|{\epsilon}_i| < E_Z/2$ are
spin polarized with spin up ($\uparrow$) (we measure all the energies
from the Fermi level). The interaction term in the Hamiltonian
(\ref{0D}) does not affect the spin polarized states, but mixes the
double-ocupied and empty states. Those states are separated
from each other by large gap $E_Z$. Therefore, this mixing is
perturbative, and it does not change ground state qualitatively.

Contrarily,  the spectrum of
the excitations, {\em e.g.} the tunneling DoS changes
drastically due to the interaction. 
Consider a spin-down electron with energy $0 < E < E_Z/2$
entering the grain. The orbital energy of this electron
${\epsilon}_0$ is close to ${\epsilon}_0  = E -E_Z/2$, and
this orbital is ocupied by a spin-up
electron. Therefore, the tunneling event creates the spin-singlet state
with the energy $2{\epsilon}_0$, and this state can mix with all the
empty states $\epsilon_i$. 
This mixing turns out to be resonant at some energy $E=E^\ast$; for
$E$ close to $E^\ast$ it requires nonperturbative treatment.

Before turning to the rigorous calculations, let us discuss this
effect qualitatively using the following simplification.  Instead of
the whole many-body system, we consider two-electrons only, however,
the single-electron orbitals for this pair are restricted by the
orbitals with ${\epsilon}_i > E_Z$ and by one orbital ${\epsilon}_0$.
The role of the rest of the electrons is to restrict the Hilbert space
for a given electron pair. This simplification is similar to the
Cooper procedure\cite{Cooper57}.  The interaction in Eq.~(\ref{0D})
involves spin singlet orbitals only. Thus, the wave function of the
electron pair $\psi$ can be labeled by one orbital index and it obeys
the Schr\"{o}dinger equation $\varepsilon \psi_i = 2
{\epsilon}_i\psi_i - \lambda \bar{\delta}\sum_j\psi_j.$ The
eigenenergies $\varepsilon$ of this equation can be found
from
\begin{equation}
\frac{\bar{\delta}}{2{\epsilon_0} - \varepsilon} +
\sum_{{\epsilon}_i>E_Z/2}\frac{\bar{\delta}}{2{\epsilon}_i -
\varepsilon}
 = \frac{1}{\lambda}. 
\label{epair}
\end{equation}
For low-lying eignestates $\varepsilon < E_Z$ one can substitute the
summation in Eq.~(\ref{epair}) by integration. Given the high-energy
cut-off $\bar{\omega}$, it yields
\begin{equation}
\frac{2\bar{\delta}}{2{\epsilon}_0 - \varepsilon} +
\ln\left(\frac{{\Delta_b}}{E_Z -
\varepsilon}\right)
 = 0, \ \ \Delta_b = \bar{\omega}e^{-2/\lambda}. 
\label{epair1}
\end{equation}
For small level spacing $\bar{\delta} \ll E_Z,
\Delta_b$,  the pair energy $\varepsilon$ is
\begin{equation}
\varepsilon\! = \!{\epsilon}_0\! +\! \frac{{\Omega}^b}{2}
\!\pm\! \left[\left(\frac{{\Omega}^b}{2}\! - \!{\epsilon}_0\right)^2\!\!+\! 
2\bar{\delta}\Delta_b
\right]^{1/2}\!\!\!\!\!\!,\ \ {\Omega}^b\! =\! E_Z\! -\! \Delta_b.
\label{wrongomega}
\end{equation}

Prior the spin down electron tunnels in, the energy of the spin-up
electron on the orbital $\epsilon_0$ was equal to
$E_\downarrow={\epsilon}_0-E_Z/2$, and thus the energy of  one electron
excitation $E = \varepsilon- E_\downarrow$ is given by
\begin{equation}
E\! = \!{E}^\ast_b
\pm \left[\left(\frac{{\Omega}^b}{2} - {\epsilon}_0\right)^2\!+ 
2\bar{\delta}\Delta_b
\right]^{1/2}\!\!\!\!\!\!, \ \ {E}^\ast_b\!=\frac{E_Z+{\Omega}^b}{2}.
\label{oneparticle}
\end{equation}
The origin of tunneling anomaly is now transparent from
Eq.~(\ref{oneparticle}): due to the repulsion between the state formed
immediately after an electron tunnels into the system and the bound
states of the Cooper pair, {\em there is no spin-down one electron
excitations } in the energy strip $|E-{E}^\ast_b|<
\left(2\bar{\delta}\Delta_b\right)^{1/2}$ -- hard gap in the DoS is
formed. It is important to emphasize that (i) the width of this gap
$\sqrt{8\bar{\delta}\Delta_b}$ significantly exceeds the single
electron level spacing and (ii) this singularity persists even when
the system is deep in the paramagnetic phase $E_Z \gg
\sqrt{2}\Delta$.

It is also noteworthy that if a spin-up electron tunnels into the grain,
it never finds the pair for itself, and, therefore, no tunneling anomaly
happens in this case. It means that the overall DoS
does not vanish but rather shows the suppression by a factor of
two. However, for the spin-polarized electrons
tunneling into the grain, we predict the complete suppression of the
tunneling DoS.   

The same arguments allow to justify the similar singularity, when
spin-up electron with energy $-E_Z/2 < E < 0$ tunnels out from the
system, while the spin down electrons tunneling from the system are
not affected.

The qualitative consideration above grasps the correct physics,
however it fails to describe the effect quantitatively, it predicts
correctly neither the position [compare ${E}^\ast_b$,
Eq.~(\ref{oneparticle}), with Eq.~(\ref{position})] nor the width of
the gap.  This is similar to the discrepancy between the binding energy
$\Delta_b$ in the original Cooper procedure and the correct BCS gap $\Delta$:
all the electrons below the Fermi energy were frozen. To remedy this
drawback, we employ a parametrically exact procedure
described now.

We start with the propagator of the superconducting fluctuations. 
The diagrammatic equation, Fig.~\ref{fig1}a,  yields
\begin{equation}
\Lambda_c(\omega) = {2\bar{\delta}}
\left[\ln\left(\frac{E_Z^2-\omega^2_+}{\Delta^2}\right)\right]^{-1}.
\label{scpresult}
\end{equation}
Here $\Delta=\bar{\omega}e^{-1/\lambda}$ is the BCS gap, and
$\omega_+=\omega+i0{\rm sgn}\omega$. 
Propagator
(\ref{scpresult}) has the pole at  $\omega = \pm \Omega$:
\begin{equation}
\Omega = \sqrt{E_Z^2-\Delta^2}.
\label{rightomega}
\end{equation}
$\Omega$ has the meaning of the bound state energy of two
quasiparticles. Using Eq.~(\ref{rightomega}) instead of oversimplified
Eq.~(\ref{wrongomega}) in the expression for ${E}^\ast_b$ in
Eq.~(\ref{oneparticle}), we obtain Eq.~(\ref{position}) for the
position of the singularity $E^\ast$.

Consider now the shape of the tunneling DoS $\nu (\omega)$ near the
singularity.  DoS for spin polarizations $\sigma=\uparrow,\downarrow$
can be expressed through the one particle Green function (GF)
$iG_{i,\sigma}(\omega)$ at zero temperature:
$\nu_\sigma(\omega)=-\frac{1}{\pi}{\rm sgn}\omega {\rm
Im}\sum_iG_{i,\sigma}(\omega)$. In its turn, the GF is given by
$1/G_{i,\sigma}=1/G_{i,\sigma}^0-\Sigma_{i,\sigma}$, where $G^0$ is
the GF for noninteracting system, $G^0_{i,\uparrow
(\downarrow)}=\left[\omega_+ -{\epsilon}_i\pm E_Z/2\right]^{-1}$, and
$\Sigma_{i,\sigma}$ is the one-particle self-energy\cite{AGD}.

The leading part of $\Sigma_{i,\downarrow}$, diagram 
Fig.~\ref{fig1}b, is given by 
\begin{equation}
-i\Sigma_{i,\downarrow}=\int\frac{d\omega_1}{2\pi}
\Lambda_c\left(\omega +\omega_1\right)G^0_{i,\uparrow}(\omega_1).
\label{selfenergy1}
\end{equation}
The singular contribution to the integral in Eq.~(\ref{selfenergy1})
comes from the positive pole of $\Lambda_c$, which corresponds exactly
to the repulsion between the bound state of the Cooper pair with the
state formed after tunneling which we discussed earlier. From
Eq.~(\ref{selfenergy1}) we obtain for the GF
\begin{equation}
G_{i,\downarrow}\!=\!\frac{\omega + {\epsilon}_i-\frac{E_Z}{2} -\Omega}{\left(\omega_+ -
{\epsilon}_i-\frac{E_Z}{2}\right)
\left(\omega_+\!+ \epsilon_i -
\frac{E_Z}{2}-\Omega \right)-\frac{\bar{\delta}\Delta^2}{\Omega}}.
\label{Greenfunction}
\end{equation} 
Summing up over all the orbitals $\epsilon_i$, and neglecting the
fine structure of the DoS on the scale of $\bar{\delta}$, we find
\begin{equation}
\nu_{\uparrow,\downarrow}=\nu_0{\cal F}_0\left(\frac{\omega \pm
E^\ast}{W_0}\right)
,\quad {\cal F}_0(x)=\frac{\theta(x^2-1)|x|}{\sqrt{x^2-1}},
\label{0Dresult}
\end{equation}
where  $\nu_0$ is the bare density of states per one spin.
Equation (\ref{0Dresult}) describes the hard gap of
the width 
\begin{equation}
W_0=\left(\frac{\bar{\delta}\Delta^2}{\Omega}\right)^{1/2},
\label{W}
\end{equation}
positioned at $\omega=E^\ast$ given by Eq.~(\ref{position}).

{\narrowtext
\begin{figure}[h]
\vspace{0.3cm}
\epsfxsize=8.7cm
\hspace*{0.2cm}
\epsfbox{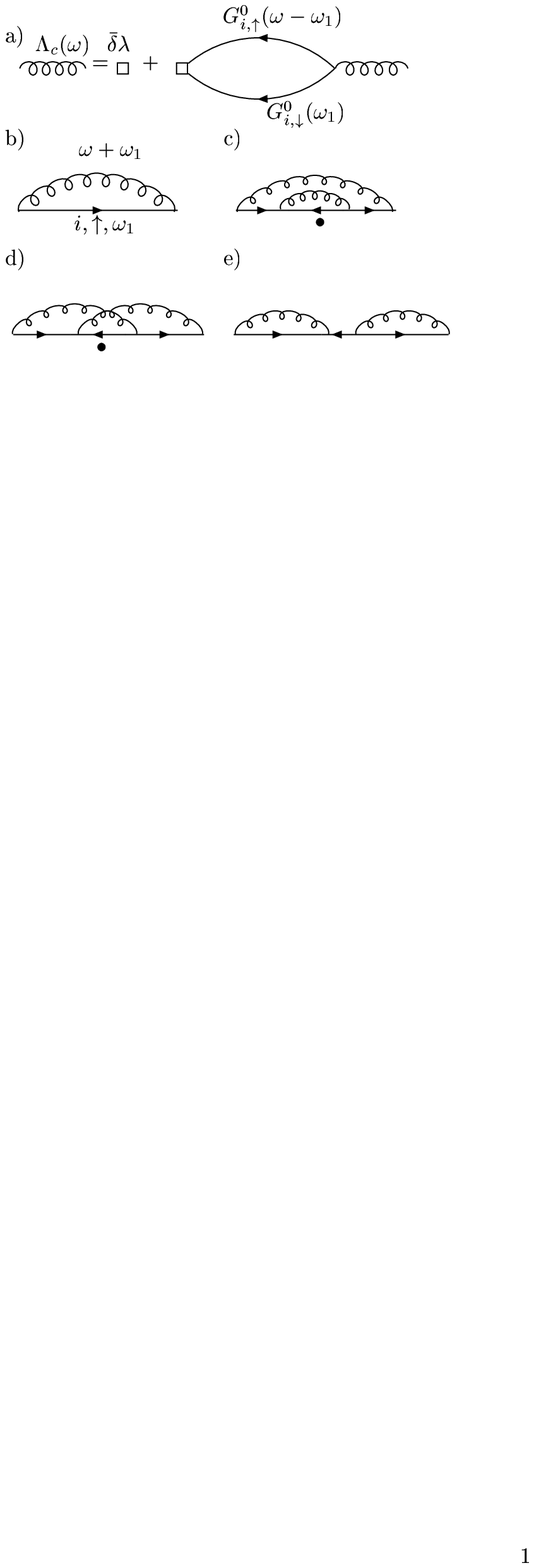}
\vspace{-5.2cm}
\caption{Diagrams describing (a) the fluctuation propagator
$\Lambda_c(\omega)$ and (b) first order contribution to the self
energy $\Sigma_{i,\downarrow}(\omega)$. 
Diagrams (c) and (d) show second order contributions
to $\Sigma$, which were neglected in comparison with reducible diagram
(e).}
\vspace{-0.26cm}
\label{fig1}
\end{figure}
}

Higher order corrections to the self-energy ({\em e.g.}  shown in
Fig.~\ref{fig1}c,d) are negligible.  Indeed, let us compare the
contributions of those diagrams with the reducible diagram
Fig.~\ref{fig1}e which is included in 
Eqs.~(\ref{selfenergy1}) and (\ref{Greenfunction}). 
 At $\omega=E^*, 2\epsilon_i =
\Omega$, all three GF in diagram (e) diverge after integration over
intermediate frequencies, whereas only two GF diverge
(non-resonant GF are indicated by $\bullet$) in
Fig.~\ref{fig1}c,d. It means that diagrams (c-d) are smaller than the
main contribution by a factor $\bar{\delta}/\Delta \ll 1$.

According to Eq.~(\ref{0Dresult}), the width $W_0$ of the hard gap is
much larger than the mean level spacing $\bar{\delta}$. However, it
apparently vanishes as $\bar{\delta}^{1/2}$ for infinite systems. To
be more precise, our $0D$ consideration was valid only if the width
of the gap $W_0\simeq\sqrt{\bar{\delta}\Delta}$ does not exceed the
Thouless energy $E_T=\hbar D/L^2$ ( $D$ is the diffusion
coefficient). Since ${\bar{\delta}} \sim 1/L^d $, ($d$ is the
dimensionality of the sample) the condition $W_0 \lesssim E_T$ breaks
down for bulk samples, $L \to \infty$, in all physical dimensions $d <
4$ and our zero-dimensional description (\ref{0D}) becomes
invalid. Nevertheless, the strong singularity
in the vicinity of $E^\ast$  persists in the tunneling DoS for
$d=1,2$ systems\cite{footnote2}.

We describe the interaction by usual Hamiltonian $\hat{\cal
H}_{int}=-\nu^{-1}_0\lambda\int dr a^\dagger_\uparrow a^\dagger_\downarrow
a_\downarrow a_\uparrow$. In the bulk system, the superconducting
fluctuations can be inhomogenous. Thus, the propagator similar to
Eq.~(\ref{scpresult}) depends on the wavevector $Q$. Diagram for such
propagator averaged over disorder is similar to  Fig.~\ref{fig1}a and
standard calculation  yields:
\begin{equation}
\Lambda_c\left(\omega,Q\right) =
{2}\left[\nu\ln\left(\frac{E_Z^2-(|\omega| +iDQ^2)^2}{\Delta^2}\right) \right]^{-1}.
\label{scpq}
\end{equation}
$\Lambda_c$ at small $Q$, ($DQ^2 \ll \Omega$) has the singularity at
$\omega$ close to the energy  $\Omega$ from
Eq.~(\ref{rightomega}). Therefore, one may expect the singularity in
DoS at the same energy $E^\ast$.

Evaluation of the first order correction $\delta\nu_\sigma^{(1)}$ to the
DoS, similar to Fig.~\ref{fig1}b,  confirms this
expectation. Integrating over intermediate frequency, averaging over
disorder, and assuming $DQ^2
\ll \Omega$, we find  the singular part of the correction\cite{elsewhere} 
\begin{equation}
\frac{\delta
\nu^{(1)}_\sigma(\omega)}{\nu_0}=\frac{\Delta^2}{2\nu_0\Omega}
{\rm Re}
\int\frac{d^dQ}{(2\pi)^d}\left[{\cal C}_\sigma\left(\omega,Q\right)\right]^2,
\label{nu1}
\end{equation}
where $d=1,2$ is the dimensionality of the sample, and 
the Cooperon ${\cal C}_\sigma$ is given by
${\cal C}_{\uparrow (\downarrow) }=\left[\omega \pm E^\ast+iDQ^2\right]$.
 For $1D$ case (strip), integration in Eq.~(\ref{nu1}) gives:
\begin{equation}
\frac{\delta
\nu^{(1)}_{\uparrow (\downarrow)}(\omega)}{\nu_0}=
\frac{{\Delta^{1/2}}}{2\nu_0\Omega D^{1/2}}
\left(\frac{\Delta}{2|\omega \pm E^\ast|}\right)^{3/2}.
\label{nu1D}
\end{equation}

In two dimensions the correction to DoS vanishes in the lowest
order.  Second order correction similar to that in Fig.~\ref{fig1}e
averaged over disorder reads\cite{elsewhere}
\[
\frac{\delta\nu^{(2)}_\sigma}{\nu_0}\!=\!
-2\!\left(\!\frac{\Delta^2}{4\nu_0\Omega}\!\right)^2\!\!
\frac{\partial}{\partial \omega}
{\rm Re}\!\!
\int\!\!\frac{d^d\!Q_1\!d^d\!Q_2}{(2\pi)^{2d}}
{\cal C}_\sigma\!\left(\omega,\! Q_1\right)
{\cal C}_\sigma^2\!\left(\omega,\! Q_2\right).
\]
For two dimensional film integration yields 
\begin{equation}
\frac{\delta\nu^{(2)}_{\uparrow(\downarrow)}(\omega)}{\nu_0}=-
\left[\frac{\Delta^2}{4g\Omega\left(\omega\pm E^\ast\right)}\right]^2
\ln\left(\frac{\Omega}{\omega\pm E^\ast }\right)^2,
\label{nu2D}
\end{equation}
where $g = 4\pi\hbar D\nu_0 \gg 1$ is the dimensionless sheet conductance of the
film in the normal state.  It is worth noticing, that the singlularity
given by Eq.~(\ref{nu2D}) is much more pronounced than that due to the
superconducting fluctuations in the normal metal\cite{Altshuler85},
$\propto g^{-1}\ln[\ln (\omega\pm E_Z)]$.

Both corrections (\ref{nu1D}) and (\ref{nu2D}) diverge with
approaching the singularity point $\omega \to \pm E^\ast$. It means that
the higher order terms have to be summed up. Here we present only the
results of this nonperturbative treatment\cite{elsewhere}. 
The resulting DoS can be
written as
\begin{equation}
\frac{\nu_{\uparrow (\downarrow)}}{\nu_0}={\cal
F}_d\left(\frac{\omega \pm E^\ast}{W_d}\right),\quad d=1,2,
\label{nuexact}
\end{equation}
where the characteristic widths 
of the singularity $W_{1,2}$ in one and two dimensons are given by
\begin{equation}
W_1=\frac{\Delta}{3}
\left(\frac{\Delta^{1/2}}{16\nu_0\Omega D^{1/2}\hbar^{1/2}}\right)^{2/3},
\quad 
W_2=\frac{\Delta^2}{4 g\Omega}.
\label{W2D}
\end{equation}
In one dimension DoS acquires a universal shape:
\begin{mathletters}
\label{shape}
\begin{equation}
{\cal F}_1(x)\!=\!1\! -\! \frac{2}{3}{\rm Re}\! \left[{1\! -\! 
i x \!\left(\!
\sqrt[3]{1\! +\! y(x)}\!+\!
\sqrt[3]{1\!-\! y(x)}
\right)
}\right]^{-1}\!\!\!\!\!,
\label{shape1}
\end{equation}
where $y(x) = \sqrt{1+ix^3}$. In two dimensions shape of the DoS
weakly depends on the conductance $g$
\begin{equation}
{\cal F}_2(x) = {\rm Re}\frac{1-z(x)}{1+z(x)}, 
\quad z=\frac{1}{-ix+\ln (4gz)}.
\label{shape2}
\end{equation}
At $x >> 1$, Eqs.~(\ref{shape}) reproduce perturbative
results  (\ref{nu1D}) and (\ref{nu2D}).  DoS
near the energy $E^\ast$ is shown in Fig.~\ref{fig2}.
\end{mathletters}

{\narrowtext
\begin{figure}[h]
\vspace{-0.4cm}
\hspace*{-1.25cm}
\epsfxsize=10.1cm
\epsfbox{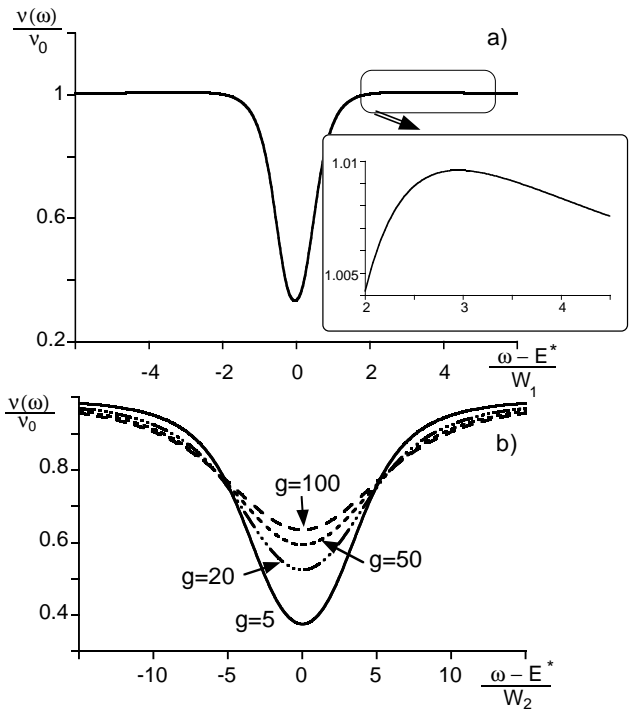}
\vspace{-2.9cm}
\caption{Singularity in DoS for spin-down polarized electrons for (a)
$1D$, and (b) $2D$ systems. Widths of the singularity $W_{1,2}$ are
given by Eq.~(\protect\ref{W2D}), and the shape is defined by
Eqs.~(\protect\ref{nuexact}) and (\protect\ref{shape}).}
\vspace{-0.26cm}
\label{fig2}
\end{figure}
}

Finally, let us mention that the DoS singularity is supressed by a
finite temperature and by the spin-orbit scattering: the singularity
remains observable only if $W \gtrsim
T,\hbar/\tau_{so}$. Corresponding expressions will be presented
elsewhere\cite{elsewhere}.

In conclusion, we have shown that the tunneling DoS of superconductors
above the paramagnetic limit has a singularity at energy determined by
Zeeman splitting and superconducting gap, Eq.~(\ref{position}). This
phenomenon is not take into account by usual BCS mean field and can
not be obtained within Gorkov-Nambu formalism. The shape of the
singularity, obtained nonperturbatively, depends on the
dimensionality of the system, Eqs.~(\ref{nuexact}) and (\ref{shape}).
We believe that our theory is adequate for existing
experiments\cite{Adams}, and suggests new effect in ultrasmall grains.

Discussions with A.I. Larkin and B.Z. Spivak are acknowledged with
gratitude.

\end{multicols} 

\begin{references}
\bibitem{footnote} This requires the size of the sytem in the
restricted direction $L$ to be parametrically smaller than the
superconducting coherence length $\xi$, $L \lesssim
\xi/(\hbar^{-1}Dm)$, where $D$ is the diffusion coefficient and $m$ is
the electron mass.
\bibitem{footnote2}
Consideration of three dimensional bulk samples is
redundant because in such system the orbital mechanism is always
dominant in suppression of superconductivity.
\end{references}
\end{document}